\documentclass[twoside]{article}
\usepackage{fleqn,espcrc2,epsf}

\newcommand{\beeq}{\begin{equation}}
\newcommand{\eneq}{\end{equation}}
\newcommand{\beeqa}{\begin{eqnarray*}}
\newcommand{\eneqa}{\end{eqnarray*}}

\usepackage{graphicx}
\usepackage[figuresright]{rotating}
\usepackage{epsfig}

\newcommand{\AmS}{{\protect\the\textfont2
  A\kern-.1667em\lower.5ex\hbox{M}\kern-.125emS}}

\title{
{\vspace{-4em} \normalsize                                             
\hfill \parbox{50mm}{DESY 99-135}}\\[25mm]
\vspace{-1cm}
Low-energy features of SU(2) Yang-Mills theory with light gluinos}

\author{Alessandra Feo\address{Institut for Theoretical Physics I, 
        University of M\"unster,\\
        Wilhelm-Klemm-Str. 9, D-48149 M\"unster, Germany \\ },
        Robert Kirchner\address{Deutsches Elektronen Synchrotron, DESY,\\ 
        Notkestr. 85, D-22603 Hamburg, Germany }
        \thanks{Talk given by Robert Kirchner.},
        Istvan Montvay$^{b}$,
        Anastassios Vladikas$^{b}$
        \thanks{On leave of absence from INFN-Rome2.}
\newline
\newline
 DESY-M\"unster Collaboration}      
 
\begin{document}

\begin{abstract}
We report on the latest results of the low-lying spectrum of bound states 
in SU(2) Yang-Mills
theory with light gluinos. The behavior of the disconnected 
contributions in the critical region is
analyzed. A first investigation of a three-gluino state is also discussed.

\vspace{1pc}
\end{abstract}

\maketitle

\section{Introduction}
The numerical simulation we report on aims at a better 
understanding of the non-perturbative low-energy features of supersymmetric 
gauge theories.
We concentrate on the simplest supersymmetric gauge theory, namely SU(2),$N=1$ 
super-Yang-Mills.
This model contains, in addition to the gauge field a massless 
Majorana fermion 
in the adjoint representation (called gluino).
For the theoretical motivation of this investigation see
\cite{conference1,conference2,bigPub} and references therein.

\section{Lattice formulation}
 We regularize 
the theory by the Wilson action as proposed in \cite{CurciVen}.
Supersymmetry is broken, both by the lattice regularization and the 
introduction of a mass term for the gluino.
The action contains two bare parameters: the gauge coupling ${\beta}$ and the
hopping parameter $K$ (bare gluino mass).
Supersymmetry is expected to be restored by tuning the bare parameters 
to their critical values \cite{CurciVen}.
The path-integral for Majorana fermions is a Pfaffian
\beeq
\int{[d\psi]} e^{-\frac{1}{2}\psi_{a}(CQ)_{ab}\psi_{b}}=\rm{Pf}(CQ), 
\eneq
where Q is the Wilson fermion matrix in the adjoint representation 
(see for example \cite{bigPub}), 
and C the charge conjugation matrix. The Pfaffian satisfies
\beeq
\rm{Pf}(CQ)^2=\det(CQ)=\det Q= \det (\tilde Q).
\eneq
$\tilde{Q}$ is the hermitean fermion matrix $\tilde{Q}=\gamma_5Q$ with 
doubly degenerate real eigenvalues, ($\det(Q)\geq 0$). In practice we 
have simulated with weight $\det(Q)^{\frac{1}{2}}$. This may lead to a 
sign problem. However, in \cite{bigPub} it is found that sign flips 
are rare.
 
\section{The low-lying spectrum}
\begin{figure}
\epsfig{file=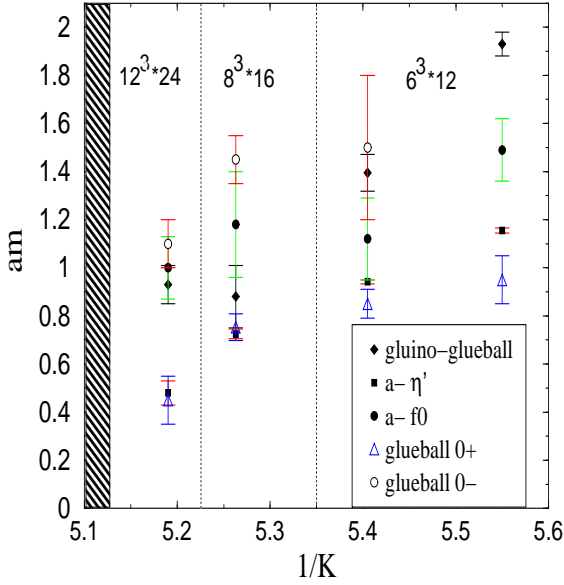, width=7.5cm,height=8cm}
\vspace{-1.5cm}
\caption{
The lightest bound state masses in lattice units as function of the bare 
gluino mass parameter $1/K$. The shaded area at $K=0.1955(5)$ is where zero 
gluino mass and supersymmetry are expected \cite{chiralSymm}.}
\vspace{-0.5cm}
\label{masses}
\end{figure}

A basic assumption about the low-energy dynamics of super-Yang-Mills theory 
is confinement, as supported by the non-vanishing string tension \cite{bigPub}.
Therefore the low-lying spectrum consists of color singlets as in QCD.
In the SUSY-limit of zero gluino mass the states should be organized in 
degenerate multiplets.
In analogy to QCD we consider scalar and pseudoscalar mesons and glueballs.
To complete the supermultiplet a spin $\frac{1}{2}$ gluino-glue
particle is also considered. 
In detail these particles and some of the 
corresponding operators are: 
\begin{itemize}
\item Scalar meson (a-f0): $ \phi_{s}=\bar{\psi}\psi$,
\item Pseudoscalar meson (a-$\eta'$): $ \phi_{p}=\bar{\psi}\gamma_{5}\psi$,
\item Gluino-glue state :$ \chi_{\alpha}=\sum_{kl}Tr(\tau_{r}U_{kl})
\psi_{\alpha}^r$,
\item $0^+$ glueball,
\item $0^-$ glueball.
\end{itemize}
For the gluino-glue state and the glueball masses blocking and smearing 
was used. The results are displayed in fig.\ref{masses}.
The presumable existence of a second multiplet requires yet another 
spin $\frac{1}{2}$ particle. 
The search for this state is an open issue.

\section{A look at the $a-\eta'$ in the critical region}
\begin{figure}
\epsfig{file=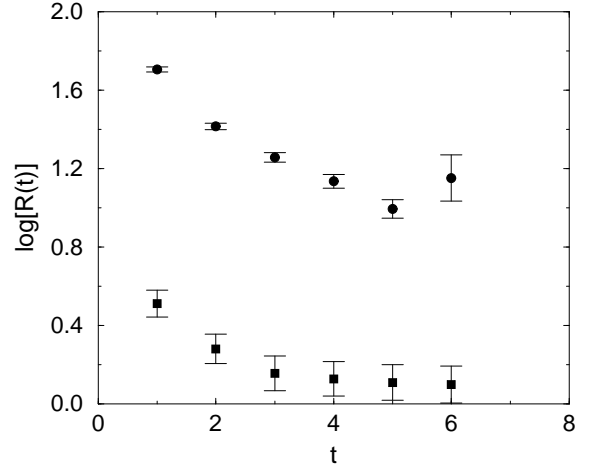, width=7.5cm}
\vspace{-1.5cm}
\caption{R(t) as defined in the text at K=0.1925 (circles) 
and K=0.196(squares)}
\vspace{-.5cm}
\label{props}
\end{figure}

The correlator of the $a-\eta'$ consists of a disconnected and a 
connected part,
\begin{eqnarray*}
C(t)=-2C(t)_{\rm{conn}}+C(t)_{\rm{disconn}}. 
\end{eqnarray*}
In QCD, $C(t)_{\rm{conn}}$
gives rise to the $\pi$-mass and $C(t)$ to the $\eta'$-mass, so that
\begin{eqnarray*}
R(t)=C(t)/C(t)_{\rm{disconn}}
\end{eqnarray*}
 is expected to decrease as we approach the chiral limit. 
In order to investigate whether this is also true in our case, 
we plot $R(t)$ in fig.\ref{props}. For $K=0.1925$ and $K=0.196$
we observe that indeed $R(t)$ demonstrates a QCD-like behavior.

\section{Investigation of a three-gluino state}
\begin{figure}
\epsfig{file=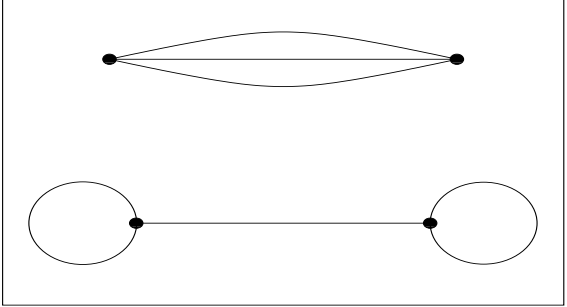, width=7.5cm, height=4.5cm}
\vspace{-0.5cm}
\caption{Contributions to the propagator of a three-gluino state. 
The second contribution arises,
since contractions of the form $\psi(x)\psi(x)$ are allowed for 
Majorana fermions.}
\label{threegluino}
\vspace{-0.5cm}
\end{figure}

Three-gluino states\footnote{We would like to thank A.Gonz\'alez-Arroyo for 
a clarifying discussion on the spin content of these particles.} can also 
be constructed in analogy to QCD baryons. This holds 
also for SU(2) since the fermions are in the adjoint representation.
In this case a possible choice for the wave function is
\beeq
\phi^{\alpha}(x)=\epsilon_{abc}(C\gamma_{4})_{\beta\gamma}\psi(x)_{a}^{\alpha}
\psi(x)_{b}^{\beta}\psi(x)_{c}^{\gamma}.
\label{wavesu2}
\eneq
This wave function which is antisymmetric in color and symmetric in spin, 
carries spin 
$\frac{3}{2}$.
\newline
For SU(3) color additional possibilities are obtained by a symmetric color 
coupling
\begin{eqnarray*}
\phi'^{\alpha}(x)=d_{abc} (C\gamma_{5})_{\beta\gamma} \psi(x)_{a}^{\alpha}
\psi(x)_{b}^{\beta}\psi(x)_{c}^{\gamma}, \\
\phi''^{\alpha}(x)=d_{abc} (C)_{\beta\gamma} \psi(x)_{a}^{\alpha}
\psi(x)_{b}^{\beta}\psi(x)_{c}^{\gamma}.
\end{eqnarray*}
The propagator of such a state has basically two contributions 
displayed in fig.\ref{threegluino}.
The correlation function $< \bar{\phi^{\alpha}}\phi^{\alpha}>$ for 
the wave function eq.(\ref{wavesu2}) has the following form:
\begin{eqnarray*}
\lefteqn{C(x,y) = - \epsilon_{a'b'c'}\epsilon_{abc}
(C\gamma_{4})_{\beta'\gamma'}(C\gamma_4)_{\beta\gamma}*  } \\
\lefteqn{\left \{ \right. 2 \Delta_{xa\alpha}^{ya'\alpha'}\Delta_{xb\beta}^{yb'\beta'} 
        \Delta_{xc\gamma}^{yc'\gamma'}+4\Delta_{xa\alpha}^{yb'\beta'}
        \Delta_{xb\beta}^{yc'\gamma'}\Delta_{xc\gamma}^{ya'\alpha'}} \\
\lefteqn { +2\Delta_{xa\alpha}^{xb\beta}\Delta_{xc\delta}^{ya'\alpha'}
\Delta_{yc'\delta'}^{yb'\beta'}C_{\gamma\delta}C_{\delta'\gamma'}
+4\Delta_{xa\alpha}^{xb\beta}\Delta_{yb'\beta'}^{xc\gamma}
\Delta_{yc'\gamma'}^{ya'\alpha'} }\\
\lefteqn{+\Delta_{xa\alpha}^{ya'\alpha'}\Delta_{xb\beta}^{xc\delta}
     \Delta_{yc'\delta'}^{yb'\beta'}C_{\gamma\delta}C_{\delta'\gamma'}} \\
\lefteqn{+2\Delta_{xa\alpha}^{yc'\delta'}\Delta_{xb\beta}^{xc\delta}
     \Delta_{yb'\beta'}^{ya'\alpha'}C_{\gamma\delta}C_{\delta'\gamma'}
\left \} \right. ,}\\ 
\end{eqnarray*}
where $\Delta = Q^{-1}$ is the gluino propagator.
The last four terms pertaining to the second ``spectacles'' graph can be 
evaluated by ``gauge averaging'' in analogy to the volume source 
method \cite{volumesource}.

\subsection{Evaluation of the spectacles graph}
We now show how to evaluate the second graph of fig.\ref{threegluino}.
With $\Omega_x$ the gauge transformation in the fundamental representation, we see that the gauge transformation in the adjoint, defined as 
$G_{x,ab}(\Omega)=[G_{x,ab}^{-1}]^T=
2Tr[\tau_{a}\Omega^{-1}(x)\tau_{b}\Omega(x)]$, obeys
\begin{eqnarray}
\lefteqn{\int d\Omega G_{a_1 b_1} =0,} \nonumber \\
\lefteqn{\int d\Omega G_{a_1 b_1}G_{a_2 b_2} G_{a_3 b_3}=
\frac{1}{6}\epsilon_{a_1 a_2 a_3}\epsilon_{b_1 b_2 b_3}.}
\label{group}  
\end{eqnarray}
The propagator $\Delta$ transforms under a gauge transformation as
\begin{eqnarray}
\Delta_{xa}^{yb} \rightarrow G^{-1}_{x, aa'}\Delta_{xa'}^{yb'}G_{y, b'b}.
\label{transf}
\end{eqnarray}
These are the necessary ingredients for an evaluation of the second graph.
We have to calculate for example (spinor indices are left out for simplicity)
\begin{eqnarray*}
\tilde{C}(x,y)\equiv
\Delta_{yc'}^{ya'}\Delta_{yb'}^{xc}\Delta_{xa}^{xb}\epsilon_{abc}
\epsilon_{a'b'c'}.
\end{eqnarray*}
First we compute the vector
\begin{eqnarray*}
W_{zb',x}=\Delta_{zb'}^{xc}\Delta_{xa}^{xb}\epsilon_{abc},
\end{eqnarray*}
for a fixed site $x$ and all sites $z$.
Next we observe that, with the help of eqs.(\ref{group}) and (\ref{transf}),
we find the identity
\begin{eqnarray*}
<\Delta_{yc'}^{za'}W_{zb',x}>=\frac{1}{6}\delta_{zy}\epsilon_{a'b'c'}
\epsilon_{abc}<\Delta_{yc}^{ya}W_{yb,x}>. 
\end{eqnarray*}
Composing now the ``shifted'' vector $W_{xc,y}^{\rm{shifted}}$,
\begin{eqnarray*}
W_{zb',x}^{\rm{shifted}}=W_{zb'-1,x}-W_{zb'+1,x}
\end{eqnarray*}
(with $W_{x4,y}=W_{x1,y}$, $W_{x0,y}=W_{x3,y}$) it can be shown that 
\begin{eqnarray*}
\sum_{y,c',b'} <\Delta_{yc'}^{zb'}W^{\rm{shifted}}_{zb',x}> = <\tilde{C}(x,y)>.
\end{eqnarray*}
To evaluate the l.h.s. of this relation numerically only one additional 
inversion is needed with $W^{\rm{shifted}}_{zb',x}$ as the source.
In this way $\left < \tilde{C}(x,y)\right >$ is obtained from a given $x$ 
to all $y$ by 
two inversions of the fermion matrix $Q$. 
An analysis of the mass of the particle characterized by eq.(\ref{wavesu2}) is
currently under way.
\newline
\newline
{\bf Acknowledgements:} 
 R. Kirchner acknowledges the financial contribution of the European
 Commission under the TMR-Program ERBFMRX-CT97-0122.



\end{document}